\documentclass[aps,twocolumn,prb]{revtex4-1}
\usepackage{graphicx}
\usepackage{amsmath}

\draft 

\begin{document}

\title{Anderson localization and delocalization of massless two-dimensional Dirac
electrons in random one-dimensional
scalar and vector potentials}

\author{Seulong Kim}
\affiliation{Department of Energy Systems Research and Department of Physics, Ajou University, Suwon 16499, Korea}
\author{Kihong Kim}
\email{khkim@ajou.ac.kr}
\affiliation{Department of Energy Systems Research and Department of Physics, Ajou University, Suwon 16499, Korea}
\affiliation{School of Physics, Korea Institute for Advanced Study, Seoul 02455, Korea}

\begin{abstract}
We study Anderson localization of massless Dirac electrons in two dimensions in one-dimensional random scalar and vector potentials theoretically for two different cases, in which the scalar and vector potentials are either uncorrelated or correlated. From the Dirac equation, we deduce the effective wave impedance, using which we derive the condition for total transmission and those for delocalization in our random models analytically. Based on the invariant imbedding theory, we also develop a numerical method to calculate the localization length exactly for arbitrary strengths of disorder. In addition, we derive analytical expressions for the localization length, which are extremely accurate
 in the weak and strong disorder limits. In the presence of both scalar and vector potentials, the conditions for total transmission and complete delocalization are generalized from the usual Klein tunneling case. We find that the incident angles at which electron waves are
 either completely transmitted or delocalized can be tuned to arbitrary values. When the strength of scalar potential disorder increases to infinity, the localization length also increases to infinity, both in uncorrelated and correlated cases.
 The detailed dependencies of the localization length on incident angle, disorder strength and energy are elucidated
 and the discrepancies with previous studies
 and some new results are discussed. All the results are explained intuitively using the concept of wave impedance.
\end{abstract}


\maketitle 

\section{Introduction}

The phenomenon of Anderson localization of quantum particles in a random potential and classical waves in random media
continues to attract a strong interest of researchers, despite of a half-century
of investigation.\cite{pwa,aalr,john,she,bk,sch,billy,mod,segev,sperl,shein}
Though the basic origin of Anderson localization, which is
the interference of wave components multiply scattered by randomly placed scattering centers, is well-understood,
new and surprising aspects of the phenomenon have been discovered
continually.\cite{izra,blio1,bpn1,nara,arn,bpn2,wchoi,king,fang}
Differences in the types of wave equations, the material properties
and the nature of disorder all affect Anderson localization strongly and can cause conceptually new phenomena to occur.
Recently, much interest has been paid to the localization arising in new kinds of condensed-matter materials and
artificially fabricated metamaterials.\cite{fang,gre,asa,hsi,liao,bab,chois,makh}

In this paper, we are especially interested in the unique localization and delocalization phenomena
occurring in the system of pseudospin-1/2 Dirac fermions in two-dimensional (2D) materials
such as monolayer graphene, which is characterized by a linear dispersion relation around the Dirac point
and described by a relativistic Dirac-type equation.\cite{kats,neto,rozh}
Though our method can be generalized to the cases of massive pseudospin-1/2 and pseudospin-1 Dirac systems, we restrict
our interest here to the massless pseudospin-1/2 case.
Our result can also be applied to the localization of quantum particles in other systems,
including cold atoms, trapped ions and semiconductors, and to that of classical electromagnetic waves in photonic systems
analogous to graphene or other 2D materials.\cite{leek,gar,zawa,deng}

The localization of aforementioned Dirac fermions in 2D in a one-dimensional (1D) random scalar or vector potential
has been studied previously by several authors.\cite{fang,zhu,bli,zhao}
In all of these studies, they considered superlattice models with random arrays
of rectangular potential barriers with random heights and widths and calculated the localization length
or the transmittance using the transfer matrix method.
One of the outstanding features of these studies is that massless Dirac particles incident normally on a 1D random scalar potential
are always delocalized regardless of the potential strength and the particle energy, whereas those incident obliquely are localized
except for in some special cases.
This phenomenon is an extension of the famous Klein tunneling
to random cases.\cite{klein,kats2}
Another interesting and counterintuitive feature is that localization is destroyed as the strength of the scalar potential disorder
increases to large values.\cite{fang}

The localization behavior of Dirac particles incident normally on a 1D random scalar potential was studied numerically
for the first time in Ref.~35,
where it has been demonstrated that, in the massless case, the localization length diverges and, in the massive case, it is
larger than the corresponding nonrelativistic value.
The transmission of electron waves through disordered Graphene superlattices with a random scalar potential was
studied in Ref.~36.
The transmittance of a finite strip was shown to be equal to 1 at normal incidence
and to have a strong angle dependence at oblique incidence.
In Ref.~37, analytical formulas for the localization length in disordered graphene superlattices
with a random scalar or vector potential
were obtained using the transfer matrix method and the weak-disorder expansion.
The localization length was reported to have a $\tan^{-2}\theta$ dependence on the incident angle $\theta$ for all energy
in a weak scalar potential and to have a $\cos^2\theta$ dependence for all energy in a weak vector potential.
More recently, the localization behavior of pseudospin-1 and pseudospin-1/2 Dirac particles
in disordered superlattices with a random scalar potential was studied in Ref.~20.
The localization length was reported to have a $\sin^{-2}\theta$ dependence for all energy and to increase to large values
in the strong disorder regime.

In this paper, we consider
the continuum Dirac equation in 2D with 1D random scalar and vector potentials, which are characterized by $\delta$-function
correlations. In addition to the model where the scalar and vector potentials are independent,
we also consider the one where
they are strongly correlated. All of our numerical results are essentially exact for all strengths of disorder.
Our main focus is to obtain the generalized condition for total transmission and those for complete delocalization in our random models
and to understand them in an intuitive and physical way.

We study our problem using three different approaches.
Starting from the Dirac equation, we deduce an analytical expression for the effective wave impedance, using which
we derive the condition for total transmission and those for delocalization in our random models analytically.
Based on the invariant imbedding method (IIM) for solving wave equations,\cite{kly,kim1,kim2,kim3,kim4,kim5} we also
derive the invariant imbedding equations for the reflection and transmission coefficients.
By applying a stochastic averaging technique to them and solving the resulting equations numerically,
we calculate the localization length exactly for an arbitrary strength of disorder.
In addition, by applying the perturbation expansion method to the invariant imbedding equations,
we derive concise analytical expressions for the localization length,
which are extremely accurate in the weak and strong disorder regimes.

One of the main advantages of the IIM for the study of localization is that it is possible to perform the disorder averaging
analytically in an exact manner and convert the random problem to an equivalent deterministic one.
In other studies, disorder was introduced in the superlattice model by assigning
the values of the potential and the layer widths randomly.
The disorder averages of various quantities were obtained by repeating a large number of calculations for many random
configurations and averaging over the results.
In order to obtain reliable disorder averages using this type of method, one usually needs to do the calculations
for a very large number of configurations.

Using our approaches, we calculate the dependencies
of the localization length on incident angle, disorder strength and particle energy in detail
and compare the results with those of the previous works. We find some crucial discrepancies and surprising new results.
We derive the incident angles at which obliquely incident electron waves are either totally transmitted or completely delocalized
for two different random models. We find that these conditions, which include the ordinary Klein tunneling as a special case, can
be understood completely through the concept of wave impedance. In the presence of a vector potential,
these angles can be tuned to arbitrary values. We also explain the counterintuitive phenomenon that, in certain cases, the localization is destroyed as the strength of scalar potential disorder increases to infinity using the impedance concept.

The rest of this paper is organized as follows. In Sec.~\ref{sec:model}, we introduce the two random models used in this study.
In Sec.~\ref{sec:met}, the IIM for the calculation of the localization length is described and the invariant
imbedding equations are derived.
In Sec.~\ref{sec:impedance}, we deduce an expression for the wave impedance, using which
we derive the condition for total transmission and those for delocalization analytically.
By applying the perturbation expansion method to the invariant imbedding equations in Sec.~\ref{sec:form}, we derive
analytical expressions for the localization length in the weak and strong disorder regimes.
In Sec.~\ref{sec:res}, we present detailed numerical results obtained using the IIM and discuss the dependencies of the
localization length on incident angle, disorder strength and particle energy.
We conclude the paper in Sec.~\ref{sec:conc} with some remarks on the implications of our results for experiments.

\section{Model}
\label{sec:model}

We are interested in the localization of massless Dirac electrons in 2D systems such as monolayer graphene
in the presence of 1D random scalar and vector potentials. Our theory can also be applied to the localization
of electromagnetic waves in graphene-like photonic systems.
The graphene layer is in the $xy$ plane and the scalar potential $U$ and the vector potential $\bf A$ ($=A_y{\hat {\bf y}}$) are
assumed to be functions of $x$ only.
The scalar potential can be generated by various methods including the electric field effect and chemical doping.
The vector potential is related to an external magnetic field
perpendicular to the graphene plane, ${\bf B}=B_0{\hat {\bf z}}$, by $B_0(x)=dA_y(x)/dx$.
It can also be induced by physical strain.

The motion of electrons in our system is described by
the 2D Dirac Hamiltonian of the $2\times 2 $ matrix form
\begin{eqnarray}
{\mathcal H}=\begin{pmatrix} U & v_F\left(\pi_x-i\pi_y\right) \\
v_F\left(\pi_x+i\pi_y\right) & U \end{pmatrix},
\end{eqnarray}
where $v_F$ ($\approx 10^6~{\rm m/s}$) is the graphene Fermi
velocity. The $x$ and $y$ components of of the kinetic momentum operator, $\pi_x$ and $\pi_y$, are given by
\begin{eqnarray}
\pi_x=\frac{\hbar}{i}\frac{d}{dx},~\pi_y=\hbar q+eA_y,
\end{eqnarray}
where
$e$ is the elementary charge and
$q$ is the $y$ component of the wave vector, which is a constant
of the motion.
The 2D stationary Dirac equation, which follows from the Hamiltonian, takes the form
\begin{eqnarray}
\frac{d}{dx}\begin{pmatrix}\psi_A \\ \psi_B \end{pmatrix}
&&=\begin{pmatrix} ka+q & ik(1-u) \\ ik(1-u) & -ka-q \end{pmatrix}\begin{pmatrix}\psi_A \\ \psi_B \end{pmatrix}\nonumber\\
&&=k\begin{pmatrix} \beta & i\epsilon \\ i\epsilon & -\beta \end{pmatrix}\begin{pmatrix}\psi_A \\ \psi_B \end{pmatrix},
\label{eq:dirac}
\end{eqnarray}
where $k$ [$=E/(\hbar v_F)$] is the wave number for the electron wave in free space
and $E$ ($>0$) is the electron energy. The dimensionless scalar and vector potentials $u$ and $a$ and the supplementary functions $\epsilon$ and $\beta$ are defined by
\begin{eqnarray}
u=\frac{U}{E},~a=\frac{eA_y}{\hbar  k},~\epsilon=1-u,~\beta=\frac{q}{k}+a.
\end{eqnarray}
It is easy to see that the localization behavior in the negative energy case is the same as that in the positive energy case with the opposite signs of the
scalar and vector potentials.

We are interested in the situation where $U$ and $A_y$ are random functions of $x$ in the region $0\le x\le L$
and zero elsewhere
and consider two different random models. In Model I,
we assume that
$u$ and $a$ are {\it independent} random functions
of $x$ and are given by
\begin{eqnarray}
u=u_0+\delta u(x),~
a=a_0+\delta a(x),
\end{eqnarray}
where $u_0$ and $a_0$ are the disorder-averaged values of $u$ and $a$
and $\delta u(x)$ and $\delta a(x)$ are Gaussian random functions satisfying
\begin{eqnarray}
&&\langle\delta u(x)\delta u(x^\prime)\rangle={\tilde g_u}\delta(x-x^\prime),~~\langle\delta u(x)\rangle=0,\nonumber\\
&&\langle\delta a(x)\delta a(x^\prime)\rangle={\tilde g_a}\delta(x-x^\prime),~~\langle\delta a(x)\rangle=0.
\end{eqnarray}
The notation $\langle\cdots\rangle$ denotes averaging over disorder and $\tilde g_u$ and $\tilde g_a$ are
independent parameters characterizing the strength of disorder.
In Model II, we consider the situation where the random functions
$\delta u(x)$ and $\delta a(x)$ are not independent, but proportional to each other such that
\begin{eqnarray}
\delta a(x)=f\delta u(x),
\end{eqnarray}
where $f$ is a real constant.
We point out that our vector potential configuration corresponds to the external magnetic field
given by
\begin{equation}
B_0(x)=\frac{\hbar k}{e}\left[a_0\delta(x)-a_0\delta(x-L)\right]+\delta B(x),
\end{equation}
where $\delta B(x)$ is a random magnetic field fluctuating between positive and negative values with zero average.
When $a_0$ is zero, we have a purely random magnetic field.
A similar type of nonrandom models with two $\delta$ functions of opposite signs
have been studied previously.\cite{agr}
From the definition of the dimensionless parameter $g_a$ to be defined in Eq.~(\ref{eq:ga2}),
we can relate it to the magnitude of the randomly-fluctuating part
of the magnetic field, $\vert \delta B\vert$, by
\begin{eqnarray}
\vert \delta B\vert\sim \sqrt{\frac{g_a \hbar E}{{l_c}^3 e^2 v_F}},
\end{eqnarray}
where $l_c$ is the correlation length of the vector potential disorder. If we substitute $E=20$ meV, $l_c=10$ nm and $g_a=0.01$,
we obtain $\vert \delta B\vert\sim 0.36$ T.

\section{Invariant imbedding method}
\label{sec:met}

We use the IIM to solve the Dirac equation in the presence of random potentials.
Similar methods have been applied in previous studies of the localization of electromagnetic waves
in random dielectric media.\cite{kim1,lee,kim6,kim7}
We assume that an electron plane wave is incident from a uniform region ($x>L$) onto the random region
($0\le x\le L$) obliquely at an angle $\theta$ and transmitted to
another uniform region ($x<0$). The wave function $\psi_A$ (or equivalently $\psi_B$) in the incident and transmitted regions can be expressed
in terms of the reflection coefficient $r$ and the transmission coefficient $t$:
\begin{eqnarray}
\psi_A\left(x,L\right)=\begin{cases}
  e^{-ip\left(x-L\right)}+r(L)e^{ip\left(x-L\right)}, & x>L \\
  t(L)e^{-ipx}, & x<0
  \end{cases},
\end{eqnarray}
where $r$ and $t$ are regarded as functions of $L$. The $y$ and (negative) $x$ components of the wave vector, $q$ and $p$, are
related to $\theta$ by $q=k\sin\theta$ and $p=k\cos\theta$. Once $\psi_A$ is obtained from the IIM,
$\psi_B$ can be calculated using
\begin{eqnarray}
\psi_B=-\frac{i}{k\epsilon}\frac{d\psi_A}{dx}+i\frac{\beta}{\epsilon}\psi_A.
\end{eqnarray}

Starting from Eq.~(\ref{eq:dirac}),
we are able to derive {\it exact} differential equations satisfied by $\mathit{r}$ and $\mathit{t}$ using the IIM
developed in Ref.~45,
which have the forms
\begin{eqnarray}
&&\frac{1}{k}\frac{dr}{dl}=-e^{-i\theta}(\epsilon\tan{\theta}-\beta\sec{\theta})+2i(\epsilon\sec{\theta}-\beta\tan{\theta})r
\nonumber\\ && ~~~~~~~~~+e^{i\theta}(\epsilon\tan{\theta}-\beta\sec{\theta})r^2,\nonumber\\
&&\frac{1}{k}\frac{dt}{dl}=i(\epsilon\sec{\theta}-\beta\tan{\theta})t+e^{i\theta}(\epsilon\tan{\theta}-\beta\sec{\theta})rt.
\label{eq:rt}
\end{eqnarray}
We can obtain $r$ and $t$ by integrating these equations
numerically from $l=0$ to $l=L$ with the initial conditions $r(0)=0$ and $t(0)=1$.

If the potentials are nonzero in the transmitted region where $x<0$, we need to solve the same differential equations
with the modified initial conditions obtained using the Fresnel formulas, which take the form
\begin{eqnarray}
r(0)= \frac{e^{-i\theta}-Q}{e^{i\theta}+Q}, ~~
t(0)= \frac{2\cos\theta}{e^{i\theta}+Q},
\label{eq:refc}
\end{eqnarray}
where $Q$ is defined by
\begin{eqnarray}
&&Q=\frac{p_2}{k\epsilon_2}-i\frac{\beta_2}{\epsilon_2},\nonumber\\
&&\epsilon_2=1-u_2,~~\beta_2=\sin\theta+a_2.
\end{eqnarray}
The parameters $u_2$, $a_2$, $\epsilon_2$ and $\beta_2$ are respectively the values of
$u$, $a$, $\epsilon$ and $\beta$ in the region $x<0$
and $p_2$ is the negative $x$ component of the wave vector in the same region.
Since $p_2$ satisfies ${p_2}^2+k^2{\beta_2}^2=k^2{\epsilon_2}^2$, we can calculate $p_2$ from
\begin{eqnarray}
\frac{p_2}{k}=\left\{\begin{matrix} \mbox{sgn}(\epsilon_2)\sqrt{{\epsilon_2}^2-{\beta_2}^2} &\mbox{if }{\epsilon_2}^2\ge {\beta_2}^2 \\
i\sqrt{{\beta_2}^2-{\epsilon_2}^2} & \mbox{if }{\epsilon_2}^2< {\beta_2}^2 \end{matrix}.\right.
\end{eqnarray}
After obtaining $r$ and $t$, we calculate the reflectance $R$ and the transmittance $T$ from
\begin{eqnarray}
&&R=\vert r\vert^2,\nonumber\\
&&T=\left\{\begin{matrix} \frac{\mbox{sgn}(\epsilon_2)\sqrt{{\epsilon_2}^2-{\beta_2}^2}}
{\epsilon_2\cos\theta}\vert t\vert^2 &\mbox{if }{\epsilon_2}^2\ge {\beta_2}^2 \\
0 & \mbox{if }{\epsilon_2}^2< {\beta_2}^2 \end{matrix}.\right.
\end{eqnarray}
With these definitions, $R$ and $T$ satisfy the law of energy conservation $R+T=1$.

One of the main advantages of using the IIM is that the disorder averaging
can be performed analytically in an exact manner.
We use Eq.~(\ref{eq:rt})
to calculate the disorder averages of various physical quantities
consisting of $r$ and $t$ exactly. In this paper, we are mainly interested
in the localization length $\xi$ defined by
\begin{eqnarray}
\xi=-\lim_{L\rightarrow\infty}\left(\frac{L}{\langle \ln T\rangle}\right).
\end{eqnarray}
The $\it nonrandom$ differential equation satisfied by $\langle \ln
T\rangle$ can be obtained using the second of Eq.~(\ref{eq:rt}) and Novikov's formula\cite{nov}
and, in the case of Model I, takes the form
\begin{eqnarray}
-\frac{1}{k}\frac{{d\langle \ln T\rangle}}{dl}&=&C_1+{\rm Re}\left[2e^{i\theta}\left(C_0
- iC_2\right)Z_1\right.\nonumber\\&&\left. -e^{2i\theta}C_1Z_2\right],
\label{eq:iie1}
\end{eqnarray}
where $Z_n$ ($n=1,2$) is equal to $\langle r^n\rangle$ and
the parameters $C_0$, $C_1$ and $C_2$ are defined by
\begin{eqnarray}
C_0&=&u_0\tan{\theta}+a_0\sec{\theta},~
C_1=g_u\tan^2{\theta}+g_a\sec^2{\theta},\nonumber\\
C_2&=&\left(g_u+g_a\right)\sec{\theta}\tan{\theta}.
\end{eqnarray}
The $\it dimensionless$ disorder parameters $g_u$ and $g_a$,
which can take any arbitrary nonnegative real values, are defined by
\begin{equation}
g_u=\tilde g_u k,~g_a=\tilde g_a k.
\label{eq:ga2}
\end{equation}
In general, our method can be applied to the case where $u_0$, $a_0$, $g_u$ and $g_a$ are
arbitrary nonrandom functions of $x$. In this paper, we restrict to
the case where they are constants independent of $x$. In that case,
the left-hand side of Eq.~(\ref{eq:iie1}) approaches asymptotically to a constant, $1/(k\xi)$,
in the $l\rightarrow \infty$ limit. Therefore we need to calculate $Z_1$ ($=\langle r\rangle$) and
$Z_2$ ($=\langle r^2\rangle$) in the $l\rightarrow \infty$ limit to obtain the localization length.

To calculate $Z_1$ and $Z_2$ for use in Eq.~(\ref{eq:iie1}), we derive an infinite number of coupled nonrandom differential
equations satisfied by the moments $Z_{n}$, where $n$ is an arbitrary nonnegative integer,
using the first of Eq.~(\ref{eq:rt}) and Novikov's formula.
These equations turn out to take the form
\begin{eqnarray}
&&\frac{1}{k}\frac{ dZ_n}{dl}=\big[2in\left(\cos\theta-u_0\sec\theta-a_0\tan\theta\right)\nonumber\\
&&~~ +g_u n^2\left(1-3\sec^2{\theta}\right)-g_a n^2\left(1+3\tan^2{\theta}\right)\big]Z_n\nonumber\\
&&~~ -ne^{i\theta}\left[C_0-i(2n+1)C_2\right]Z_{n+1}\nonumber\\
&&~~ +ne^{-i\theta}\left[C_0-i(2n-1)C_2\right]Z_{n-1}\nonumber\\
&&~~ +\frac{1}{2}n(n+1)e^{2i\theta}C_1Z_{n+2}\nonumber\\
&&~~ +\frac{1}{2}n(n-1)e^{-2i\theta}C_1Z_{n-2}.
\label{eq:iie2}
\end{eqnarray}
If the potentials are zero in $z>L$ and $z<0$, the initial conditions for $Z_{n}$'s are $Z_{0}=1$ and
$Z_{n}(l=0)=0$ for $n>0$. In the $l\rightarrow \infty$ limit,
all $Z_{n}$'s become independent of $l$ and
the left-hand sides of these equations vanish. Then we obtain an infinite
number of coupled $\it algebraic$ equations.
The moments
$Z_{n}$ with $n > 0$ are coupled to one another
and their magnitudes decrease rapidly as $n$ increases.
Based on this observation, we solve these algebraic equations numerically by a
systematic truncation method described in Ref.~41.

In Model II, $\delta u(x)$ and $\delta a(x)$ are not independent, but proportional to each other.
This condition leads to completely different equations for $Z_{n}$ and $\langle \ln T\rangle$.
The invariant imbedding equation for $Z_{n}$ in this case is written as
\begin{eqnarray}
&&\frac{1}{k}\frac{dZ_n}{dl}=\Big\{2in\left(\cos\theta-u_0\sec\theta-a_0\tan\theta\right)\nonumber\\&&~~
\left. +g_u n^2\left[1-f^2-3\left(\sec{\theta}+f\tan{\theta}\right)^2\right]\right\}Z_n\nonumber\\
&&~~-ne^{i\theta}\left[C_0-i(2n+1)D_2\right]Z_{n+1}\nonumber\\
&&~~+ne^{-i\theta}\left[C_0-i(2n-1)D_2\right]Z_{n-1}\nonumber\\
&&~~+\frac{1}{2}n(n+1)e^{2i\theta}D_1Z_{n+2}\nonumber\\
&&~~+\frac{1}{2}n(n-1)e^{-2i\theta}D_1Z_{n-2},
\label{eq:iie3}
\end{eqnarray}
where
$D_1$ and $D_2$ are defined by
\begin{eqnarray}
D_1&=&g_u\left(\tan{\theta}+f\sec{\theta}\right)^2,\nonumber\\
D_2&=&g_u\left(\sec{\theta}+f\tan{\theta}\right)\left(\tan\theta+f\sec\theta\right).
\end{eqnarray}
The equation for the localization length takes the form
\begin{eqnarray}
&&\frac{1}{k\xi}=-\frac{1}{k}\lim_{l\rightarrow\infty}\left(\frac{{d\langle \ln T\rangle}}{dl}\right)\nonumber\\
&&~~~~=D_1+{\rm Re}\left[2e^{i\theta}\left(C_0
- iD_2\right)Z_1\left(l\rightarrow\infty\right)\right.\nonumber\\&&\left. ~~~~~~~-e^{2i\theta}D_1Z_2\left(l\rightarrow\infty\right)\right].
\label{eq:iie4}
\end{eqnarray}

\section{Interpretation of Klein tunneling and related delocalization phenomena using the concept of wave impedance}
\label{sec:impedance}

\subsection{Effective wave impedance}

Klein tunneling originally refers to the phenomenon that Dirac fermions entering a large potential barrier
are transmitted almost completely as the barrier becomes higher and wider.\cite{klein}
In the case of massless Dirac fermions such as the electrons in monolayer graphene, the transmission is perfect
regardless of the strength and the shape of the potential when electrons are incident normally on it.\cite{kats2}
Such perfect transmission occurs even in the case where the potential is a 1D random function of the position.
This unique delocalization phenomenon, which we call Klein delocalization, has been demonstrated in some previous papers.\cite{zhu,bli,zhao}

Several different concepts and approaches, which include, for example, particle-antiparticle pair production
and chirality conservation, have been used to explain
Klein tunneling and related Klein delocalization phenomena.
In this paper, we approach these phenomena from the viewpoint of classical wave propagation theory
and demonstrate that they can be interpreted readily using the concept of wave impedance.

With that purpose in mind, we rewrite the Dirac equation, Eq.~(\ref{eq:dirac}), as
the wave equation for $\psi_A$ by eliminating $\psi_B$, which takes the form
\begin{eqnarray}
\frac{d}{dx}\left(\frac{1}{\epsilon}\frac{d\psi_A}{dx}\right)+p^2\epsilon\eta^2\psi_A=0,
\end{eqnarray}
where $\eta$ is defined by
\begin{eqnarray}
\eta^2=\frac{1}{\cos^2\theta}\left[1-\left(\frac{\beta}{\epsilon}\right)^2-\frac{1}{k\epsilon}\frac{d}{dx}
\left(\frac{\beta}{\epsilon}\right)\right].
\label{eq:imp1}
\end{eqnarray}
We notice that the wave equation of this form looks the same as that for $p$-polarized electromagnetic waves propagating normally in a
medium with the effective wave impedance given by $\eta(x)$.

In the incident region where the potentials $u$ and $a$ are zero,
$\epsilon$ is equal to 1 and $\beta$ is equal to $\sin\theta$, therefore $\eta$ is equal to 1 for all $\theta$.
If $\eta$ is unity in all other parts of the space as well, then there will be no wave reflection and the transmittance will be 1.
From the form of $\eta$, we find that total transmission can arise only when the parameter $\beta/\epsilon$ is
a constant independent of $x$.
In that case, the expression for $\eta$ is simplified to
\begin{eqnarray}
\eta^2=\frac{1}{\cos^2\theta}\left[1-\left(\frac{\beta}{\epsilon}\right)^2\right].
\label{eq:imp2}
\end{eqnarray}

Sometimes, it is more convenient to rewrite Eq.~(\ref{eq:dirac}) as another alternative form
\begin{eqnarray}
\frac{d}{dx}\left(\frac{1}{\beta}\frac{d\psi_A}{dx}\right)-\frac{1}{\epsilon}\frac{d}{dx}\left(\frac{\epsilon}{\beta}\right)
\frac{d\psi_A}{dx}+p^2\beta\tilde\eta^2\psi_A=0,
\end{eqnarray}
where $\tilde\eta$ is given by
\begin{eqnarray}
{\tilde\eta}^2=\frac{1}{\cos^2\theta}\left[\left(\frac{\epsilon}{\beta}\right)^2+\frac{1}{k\epsilon}\frac{d}{dx}
\left(\frac{\epsilon}{\beta}\right)-1\right].
\label{eq:imp4}
\end{eqnarray}
When $\epsilon/\beta$ is independent of $x$, these equations are simplified to
\begin{eqnarray}
&&\frac{d}{dx}\left(\frac{1}{\beta}\frac{d\psi_A}{dx}\right)+p^2\beta\tilde\eta^2\psi_A=0,\nonumber\\
&&{\tilde\eta}^2=\frac{1}{\cos^2\theta}\left[\left(\frac{\epsilon}{\beta}\right)^2-1\right].
\label{eq:imp3}
\end{eqnarray}
Again, the uniformity of $\epsilon/\beta$ implies that of the effective wave impedance $\tilde\eta$.

\subsection{Generalized Klein tunneling in inhomogeneous potentials}
\label{sec:kl}

The condition for total transmission is obtained by setting $\eta$ equal to 1 in Eq.~(\ref{eq:imp2}).
We obtain two solutions, which are
$a=-u\sin\theta$ and $a=(u-2)\sin\theta$, respectively. Only the first of these two is the real solution
because, in the second case, $\beta/\epsilon$ changes discontinuously from $\sin\theta$ to $-\sin\theta$ as the wave
enters obliquely from the incident region and the derivative term in Eq.~(\ref{eq:imp1}) is nonzero.
Therefore the expression for the incident angle $\theta_K$ at which
the transmittance is identically equal to 1 becomes
\begin{eqnarray}
\sin\theta_K=-\frac{a}{u}=-\frac{ev_FA_y}{U}.
\label{eq:ktc}
\end{eqnarray}

If there is no vector potential, then this condition is satisfied at $\theta_K=0$
for any arbitrary functional form of $U(x)$, which corresponds to the ordinary Klein tunneling
at normal incidence. In the presence of a vector potential, $\theta_K$ exists only if $ a/u $ is
a real constant in the range $-1<a/u<1$. Then the analogy of Klein tunneling arises for particles incident at
a nonzero angle $\theta_K$. It is interesting to note that this condition can be
satisfied even if $a(x)$ and $u(x)$ are arbitrary functions of $x$, including random functions, as long as their ratio is constant.

\begin{figure}
\centering\includegraphics[width=\linewidth]{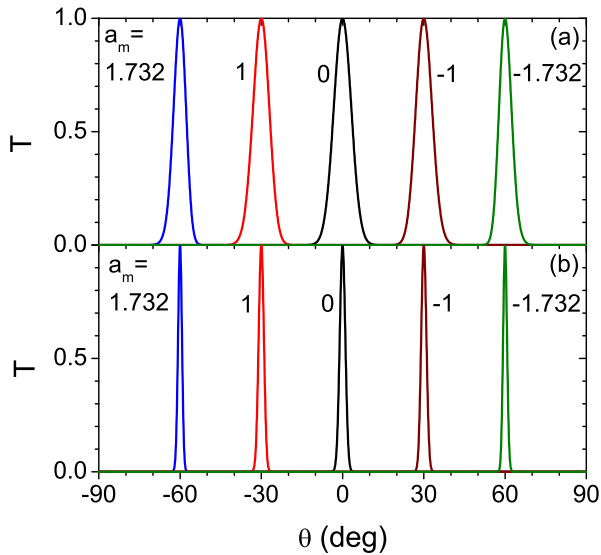}
 \caption{Transmittance $T$ plotted versus incident angle $\theta$, when massless Dirac particles
 are incident on an inhomogeneous strip of thickness $L$ with the normalized scalar potential, $u=2-2x/L$,
 and the normalized vector potential, $a=a_m(1-x/L)$ ($a_m=1.732$, 1, 0, $-1$, $-1.732$),
 in the region $0\le x\le L$.
 In the transmitted region ($x<0$), the potentials are $u=2$ and $a=a_m$.
 The strip thickness is given by (a) $kL=100$ and (b) $kL=1000$.}
 \label{fig:a}
 \end{figure}

In Fig.~\ref{fig:a}, we illustrate the total transmission phenomenon at oblique angles of incidence,
when $u$ and $a$ are given by the linear functions $u=u_m(1-x/L)$ and $a=a_m(1-x/L)$ in the region $0\le x\le L$.
In the incident region ($x>L$), $u$ and $a$ are both zero, and in the transmitted region ($x<0$), we set $u=u_m$ and $a=a_m$
to make the potentials continuous.
The linear vector potential corresponds to an external magnetic field of $\vert B_0\vert=Ea_m/(ev_FL)$
in the $-z$ direction, while the linear scalar potential corresponds to an external electric field of $\vert {\mathcal E}_0\vert=Eu_m/(eL)$
in the $-x$ direction. Since the ratio of $a$ and $u$ is a constant, we find
$\theta_K=-60^\circ$, $-30^\circ$, $0^\circ$, $30^\circ$ and $60^\circ$
when $u_m$ and $a_m$ are given by
$u_m=2$ and $a_m=1.732$, 1, 0, $-1$ and $-1.732$, which is clearly verified by the sharp transmission peaks shown in Fig.~\ref{fig:a}.
These peaks become sharper as the thickness of the nonuniform region, $L$, gets larger.
The angle $\theta_K$ can be tuned easily
by tuning either the external magnetic field or the external electric field.
We notice that our system in the present geometry can function as a very efficient directional filter.

\subsection{Delocalization condition in the presence of random scalar and vector potentials}
\label{sec:deloc}

Now we return to the case where the potentials are random.
We are interested in deriving the delocalization condition, which is in general distinct from
the total transmission condition, Eq.~(\ref{eq:ktc}).
The wave is delocalized and the localization length diverges only if the effective impedance $\eta$ is nonrandom, whereas
the total transmission occurs
when $\eta$ is identically equal to 1. Therefore the delocalization condition is a much weaker condition.

There are three different ways in which the nonrandomness or uniformity of $\beta/\epsilon$ can be achieved.
If only the scalar potential $u$ (therefore, only $\epsilon$) is random and the vector potential is constant,
then we are forced to set $\beta=0$ in order to make $\beta/\epsilon$ nonrandom.
In this case, the wave is delocalized at the
incident angle $\theta_R$ satisfying
\begin{eqnarray}
\sin\theta_R=-a_0,
\label{eq:angle1}
\end{eqnarray}
where $\vert a_0\vert<1$.
When $a_0$ is zero, this gives $\theta_R=0$,
which is the same as the total transmission condition and
corresponds to the usual Klein tunneling of normally-incident massless Dirac electrons in a 1D random scalar potential.
In general cases where $a_0$ is nonzero, however, this condition is different from Eq.~(\ref{eq:ktc})
and the two angles $\theta_R$ and $\theta_K$ are different.
Since the medium is effectively uniform when the wave is incident at $\theta_R$, it propagates without an exponential decay.
However, since the effective impedance given by $\eta=\vert\cos\theta_R\vert^{-1}$ is not equal to 1, the wave is reflected at the interfaces and the disorder-averaged transmittance
of a strip of finite thickness is smaller than 1.

Secondly, when only the vector potential $a$ (therefore, only $\beta$) is random
and the scalar potential is constant, it is more convenient to use the alternative expression Eq.~(\ref{eq:imp3}), from which
a necessary condition for the nonrandomness of $\tilde\eta$ is seen to be $\epsilon=0$, or equivalently, $E=U$. In this case, however,
the value of ${\tilde\eta}^2$ is $-1/\cos^2\theta$, and therefore the effective impedance $\tilde\eta$ is imaginary.
Then the wave becomes evanescent and decays exponentially in a manner similar to localized waves.
An exception to this occurs when $\beta_0$ ($=\sin\theta+a_0$)
is also zero simultaneously.
In later sections, we will show
analytically and numerically that the localization length in this case indeed diverges.

Finally, when both $u$ and $a$ are random, a necessary condition for $\eta$ to be nonrandom is that $\delta a(x)$ and $\delta u(x)$
have to be proportional to each other, such that $\delta a(x)=f\delta u(x)$, as in our Model II. Then the condition that
$\beta/\epsilon$ is uniform gives
\begin{eqnarray}
\sin\theta_S=-a_0-f(1-u_0),
\label{eq:angle2}
\end{eqnarray}
where the constant in the right-hand side should satisfy
$\vert a_0+f(1-u_0)\vert<1$. In this case, we define $\theta_S$ as the delocalization angle.
We also need an additional constraint $\vert f\vert <1$ to have a propagating wave.
The effective impedance at $\theta_S$ given by
\begin{eqnarray}
\eta^2=\frac{1-f^2}{\cos^2\theta_S}
\end{eqnarray}
is not equal to 1.
If $\vert f\vert<1$, $\eta$ is real and the disorder-averaged transmittance
of a finite strip is smaller than 1. The total transmission can be obtained if $a_0$ is proportional to $u_0$ with the same
proportionality constant $f$, such that $a(x)=f u(x)$. This case just corresponds to Eq.~(\ref{eq:ktc}) with both $u$ and $a$ as random functions.
On the other hand, if $\vert f\vert\ge 1$, $\eta$ is either zero or imaginary and the wave decays exponentially.

\subsection{Alternative derivation of the delocalization condition using the Fresnel formula}

Alternatively, we can derive the delocalization condition using the Fresnel formula for the reflection coefficient.
We consider our stratified random medium as consisting of a large number of
very thin strips. The reflection coefficient between two neighboring strips is written as
\begin{eqnarray}
r=\frac{p/\epsilon-p^\prime/\epsilon^\prime-ik\left(\beta/\epsilon-\beta^\prime/\epsilon^\prime\right)}
{p/\epsilon+p^\prime/\epsilon^\prime+ik\left(\beta/\epsilon-\beta^\prime/\epsilon^\prime\right)},
\end{eqnarray}
where $p$ ($p^\prime$) is the negative $x$ component of the wave vector in the first (second) strip with
the parameters $\epsilon$ and $\beta$ ($\epsilon^\prime$ and $\beta^\prime$). Since $p$ satisfies
$p^2+k^2\beta^2=k^2\epsilon^2$, we can introduce $\phi$ (and similarly $\phi^\prime$) such that
\begin{eqnarray}
\frac{p}{\epsilon}=k\cos{\phi},~~\frac{\beta}{\epsilon}=\sin{\phi}.
\end{eqnarray}
Then $r$ is written as
\begin{eqnarray}
r=\frac{e^{-i\phi}-e^{-i\phi^\prime}}{e^{i\phi}+e^{-i\phi^\prime}},
\end{eqnarray}
which vanishes when $\phi=\phi^\prime$, or equivalently, when $\beta/\epsilon=\beta^\prime/\epsilon^\prime$.
If this condition is maintained throughout the system, we expect to have perfect transmission and delocalization.

\subsection{Counterintuitive delocalization in the strong disorder limit}
\label{sec:sdl}

Our formalism based on the concept of wave impedance can be used to predict and explain the counterintuitive delocalization phenomena
arising in the infinite disorder limit in some cases.
We first consider the case where only the scalar potential is random. Then, from the expression of the impedance, Eq.~(\ref{eq:imp1}),
we find that in the strong disorder limit where $\delta u$ is statistically much larger than 1, the last two terms become negligibly small
and the impedance approaches a constant given by $\eta\approx \vert\cos\theta\vert^{-1}$. Therefore, as the disorder parameter $g_u$ approaches infinity,
the system becomes less and less random and the localization length should diverge for all $\theta$.
This behavior will be verified analytically by solving our invariant imbedding equations using
a strong-disorder expansion in Sec.~\ref{sec:sdr} and confirmed in a precise numerical calculation in Sec.~\ref{sec:dd}.

Next, we consider the case where only the vector potential is random. Then from the alternative expression of the impedance, Eq.~(\ref{eq:imp4}),
we find that in the strong disorder limit where $\delta a$ is statistically much larger than 1, the impedance approaches a constant given by ${\tilde\eta}^2\approx -1/\cos^2\theta$. Therefore the impedance is nonrandom, but imaginary, which leads to a finite decay rate
independent of the disorder strength $g_a$ in the $g_a\rightarrow\infty$ limit. The analytical expression of the localization length in the large
disorder limit to be derived in the next section will show just such a behavior.

Finally, in the case where $\delta a(x)$ and $\delta u(x)$
are proportional to each other such that $\delta a(x)=f\delta u(x)$,
we find from Eq.~(\ref{eq:imp1}) that in the strong disorder limit, the impedance approaches a constant given by $\eta^2\approx (1-f^2)/\cos^2\theta$. When $\vert f\vert$ is smaller than 1, this constant is real and we have a complete delocalization
and a diverging localization length for all $\theta$ in the large disorder limit, whereas, when $\vert f\vert\ge 1$, the impedance is either zero or imaginary.
In this latter case, the localization length approaches a constant as the disorder parameter increases to infinity. These behaviors
will be also confirmed in the next sections.

\section{Analytical expressions for the localization length in the weak and strong disorder regimes}
\label{sec:form}

\subsection{Weak disorder regime}

Starting from the invariant imbedding equations, Eqs.~(\ref{eq:iie1}), (\ref{eq:iie2}), (\ref{eq:iie3}) and (\ref{eq:iie4}),
and applying the perturbation theory, it is
possible to derive accurate analytical expressions for the localization length in the weak and strong disorder limits.
We first consider the weak disorder regime.
We write the reflection coefficient $r$ as $r=r_0+\delta r$,
where $r_0$ is the reflection coefficient from an interface between free space and a half-space medium
with the parameters $\epsilon_0$ and $\beta_0$. The expression for $r_0$ can be obtained from that of $r(0)$
in Eq.~(\ref{eq:refc}) by replacing $\epsilon_2$ and $\beta_2$ with $\epsilon_0$ and $\beta_0$ respectively.

By substituting $r=r_0+\delta r$ into the infinite number of algebraic equations obtained from Eq.~(\ref{eq:iie2}),
we get an infinite number of coupled equations for $\langle \left(\delta r\right)^n\rangle$ for all integers $n$.
We expand these averages, which are at least of the first order in $g_u$ and $g_a$, in terms of the small perturbation parameters $g_u$
and $g_a$.
From analytical considerations and numerical calculations, we can demonstrate that
the leading terms for $\langle \delta r\rangle$ and $\langle \left(\delta r\right)
^2 \rangle$ are of the first order, while
that of $\langle \left(\delta r\right)^3 \rangle$ is of the second order, except at incident angles close to the critical angle of total
reflection. From this consideration, we substitute
\begin{eqnarray}
&&Z_1=r_0+\langle \delta r\rangle,~Z_2=r_0^2+2r_0\langle \delta r\rangle+\langle \left(\delta r\right)
^2 \rangle,\nonumber\\
&&Z_3\approx r_0^3+3r_0^2\langle \delta r\rangle+3r_0\langle \left(\delta r\right)^2 \rangle,
\end{eqnarray}
into Eq.~(\ref{eq:iie2}) when $n=1$ and 2 in the $l\rightarrow\infty$ limit
and obtain two coupled equations for $\langle \delta r\rangle$ and $\langle \left(\delta r\right)^2 \rangle$.
We solve these equations analytically and substitute the resulting expression for $\langle \delta r\rangle$ into Eq.~(\ref{eq:iie1})
to the leading order in the disorder parameters. We note that to the leading order, we
have a simplification and do not need $\langle \left(\delta r\right)^2 \rangle$
in this equation. The result for the localization length in Model I is very simple and takes the form
\begin{eqnarray}
\frac{1}{k\xi}=2\sqrt{\beta_0^2-\epsilon_0^2}\Theta(\beta_0^2-\epsilon_0^2)
+\frac{g_u \beta_0^2+g_a \epsilon_0^2}{\epsilon_0^2-\beta_0^2},
\label{eq:af1}
\end{eqnarray}
where $\Theta$ is the step function, $\Theta(x)=1$ for $x>0$ and 0 for $x<0$.

From the form of Eq.~(\ref{eq:af1}), it follows that
there is a symmetry under the sign change of $\epsilon_0$ and $\beta_0$.
It turns out that this feature is not limited to the weak disorder regime, but is valid
in all parameter ranges including the intermediate and strong disorder regimes.
If only the scalar potential is random (therefore, $g_a=0)$ and $\vert\epsilon_0\vert$ is
greater than $\vert\beta_0\vert$, the explicit form of the normalized localization length is
\begin{eqnarray}
k\xi=\frac{\left(1-u_0\right)^2-\left(\sin\theta+a_0\right)^2}{g_u\left(\sin\theta+a_0\right)^2},
\end{eqnarray}
which is valid for all $\theta$.
If $u_0\ne 1$, this expression diverges at $\theta=\theta_R$ given by $\sin\theta_R=-a_0$, as expected from
our argument based on the impedance concept.
When $u_0=a_0=0$, it reduces to $k\xi=(g_u\tan^2\theta)^{-1}$, which diverges at $\theta=0$.
The same $\theta$ dependence was reported in Ref.~37, though the result there differs from ours in that the
$\tan^{-2}\theta$ dependence was obtained for both the $u_0= 0$ and $u_0\ne 0$ cases.
If $u_0\ne 1$ and $a_0=0$, $k\xi$ can be approximated as
$k\xi\approx ({\epsilon_0}^2/g_u)\sin^{-2}\theta$ near $\theta=0$.
The same $\sin^{-2}\theta$ dependence was reported in Ref.~20, though the average scalar potential corresponding to our $u_0$
was zero in that work.
In the more general cases where both $u_0$ and $a_0$ are nonzero, we can write an approximate expression near $\theta_R$ as
\begin{eqnarray}
k\xi\approx \frac{\left(1-u_0\right)^2}{\left(g_u\cos^2\theta_R\right)\left(\theta-\theta_R\right)^2}.
\label{eq:z1}
\end{eqnarray}
We have verified numerically using the IIM that the $\left(\theta-\theta_R\right)^{-2}$ dependence is not limited
to the weak disorder regime, but is valid
in all parameter ranges including the intermediate and strong disorder regimes in Model I with only scalar potential disorder.

Another interesting point to make is that in the total reflection (or tunneling)
regime where $\vert\beta_0\vert>\vert\epsilon_0\vert$, the localization length
increases as the disorder parameter increases, as can be seen easily from Eq.~(\ref{eq:af1}).
This is an example of the well-known disorder-enhanced tunneling phenomenon,\cite{kim7,frei,luck,kim_t,hein}
which has not been discussed before
in the context of Dirac electrons in a random potential.

Next, we consider the case where only the vector potential is random (therefore, $g_u=0$).
If $\epsilon_0\ne 0$ (that is, $u_0\ne 1$), the localization length
does not diverge and takes a maximum value equal to ${g_a}^{-1}$
at the angle satisfying $\beta_0=0$, away from which it decreases parabolically.
In the special case where $u_0=a_0=0$, $k\xi$ is given by $k\xi=\cos^2\theta/g_a$.
The same $\cos^2\theta$ dependence was reported in Ref.~37.
The divergent behavior can occur if $\epsilon_0$ is zero, as we have argued in Sec.~\ref{sec:deloc}. In that case, we find from Eq.~(\ref{eq:af1}) that
\begin{eqnarray}
k\xi=\frac{1}{2\vert\beta_0\vert},
\label{eq:z3}
\end{eqnarray}
which diverges as $\beta_0\to 0$.
The disorder-enhanced tunneling phenomenon also occurs in this case.

In a similar manner as in Model I, we can derive the expression for the localization length for Model II
in the weak disorder regime, which takes the form
\begin{eqnarray}
\frac{1}{k\xi}=2\sqrt{\beta_0^2-\epsilon_0^2}\Theta(\beta_0^2-\epsilon_0^2)
+\frac{g_u \left(\beta_0+f \epsilon_0\right)^2}{\epsilon_0^2-\beta_0^2}.
\label{eq:af2}
\end{eqnarray}
We observe that the symmetry with respect to the sign change of $\epsilon_0$ and $\beta_0$ is absent.
The localization length can diverge if the incident angle satisfies $\beta_0+f \epsilon_0=0$, which is the same condition
as Eq.~(\ref{eq:angle2}) obtained in Sec.~\ref{sec:deloc}. However, this is not sufficient, but we need an additional condition $\vert\epsilon_0\vert>\vert\beta_0\vert$
in order not to have the first term in the right-hand side of Eq.~(\ref{eq:af2}), which gives the additional constraint $\vert f\vert<1$.
Near the angle $\theta_S$ given by Eq.~(\ref{eq:angle2}), $k\xi$ is approximated as
\begin{eqnarray}
k\xi\approx \frac{\left(1-f\right)^2\left(1-u_0\right)^2}{\left(g_u\cos^2\theta_S\right)\left(\theta-\theta_S\right)^2}.
\end{eqnarray}
We have verified that the $\left(\theta-\theta_S\right)^{-2}$ dependence is not limited to the weak disorder regime, but is valid
in all parameter ranges in Model II with $\vert f\vert<1$.
On the other hand, the case with $\vert f\vert>1$ shows
a close similarity to Model I with only vector potential disorder.
From Eq.~(\ref{eq:af1}), it is straightforward to show that the localization length
takes a finite maximum value equal to ${[g_u(f^2-1)]}^{-1}$
at the angle satisfying $\beta_0=-\epsilon_0/f$.
In addition, the disorder-enhanced tunneling phenomenon occurs in Model II regardless of the value of $f$.

\begin{figure}
\centering\includegraphics[width=\linewidth]{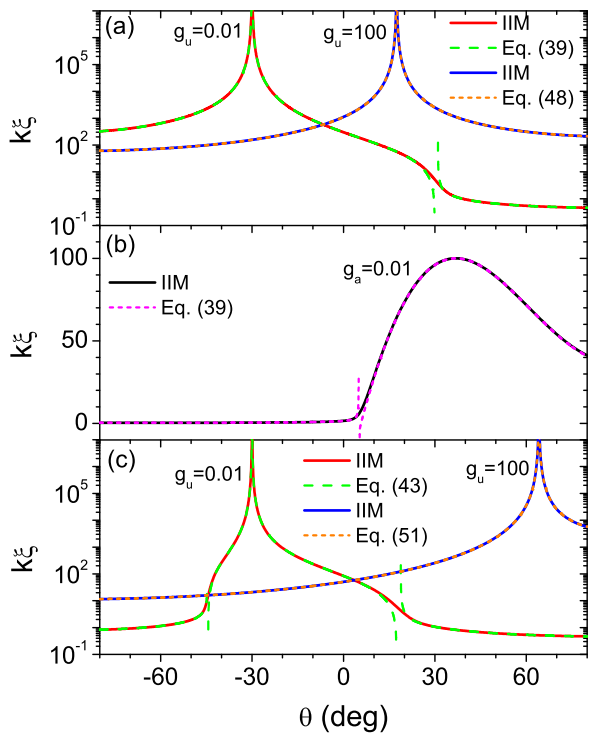}
 \caption{Illustration of the comparison between the numerical results obtained using the IIM and the analytical formulas
 for the localization length presented in Sec.~\ref{sec:form}.
 (a) Normalized localization length $k\xi$ plotted versus incident angle $\theta$ for electron waves in Model I,
 when $g_a=0$ and $u_0=0$. The left curve is for $g_u=0.01$ and $a_0=0.5$ and is compared with Eq.~(\ref{eq:af1}), while
 the right curve is for $g_u=100$ and $a_0=-0.3$ and is compared with Eq.~(\ref{eq:ssp}).
 (b) Normalized localization length versus incident angle in Model I,
 when $g_a=0.01$, $g_u=0$, $u_0=0.5$ and $a_0=-0.6$. The IIM result is compared with Eq.~(\ref{eq:af1}).
 (c) Normalized localization length versus incident angle in Model II,
 when $f=0.6$ and $u_0=0.5$. The left curve is for $g_u=0.01$ and $a_0=0.2$ and is compared with Eq.~(\ref{eq:af2}), while
 the right curve is for $g_u=100$ and $a_0=-1.2$ and is compared with Eq.~(\ref{eq:ssvp}).}
 \label{fig:comp}
 \end{figure}

We have made extensive comparisons between our exact numerical results obtained using the IIM
and the analytical formulas, Eqs.~(\ref{eq:af1}) and (\ref{eq:af2})
and found that both of these equations are extremely accurate except very close to the region where $\vert\beta_0\vert=\vert\epsilon_0\vert$, if the disorder parameters are
sufficiently small. We show some examples of this comparison in Fig.~\ref{fig:comp} to illustrate the accuracy.

\subsection{Strong disorder regime}
\label{sec:sdr}

In this subsection, we present the analytical expressions for the localization length in the strong disorder regime.
The idea of the perturbation theory is similar to the weak disorder case, but instead of expressing $r$ as $r=r_0+\delta r$,
we write $r=r_\infty+\delta r$, where $r_\infty$ is the reflection coefficient from an interface between free space and an
infinitely-disordered half-space medium
with the parameters $\epsilon_0$ and $\beta_0$,
which is given by
\begin{eqnarray}
r_\infty=\left\{\begin{matrix} \frac{e^{-i\theta}-1}{e^{i\theta}+1}, & {\mbox {if}}~\epsilon_0>0 \cr
\frac{e^{-i\theta}+1}{e^{i\theta}-1}, & {\mbox {if}}~\epsilon_0<0 \end{matrix}.\right.
\end{eqnarray}

We first consider Model I with only scalar potential disorder.
Similarly to the weak disorder case,
we substitute $r=r_\infty+\delta r$ into Eq.~(\ref{eq:iie2}) and
get an infinite number of coupled equations for $\langle \left(\delta r\right)^n\rangle$.
We expand these averages in terms of the small perturbation parameter ${g_u}^{-1}$.
From analytical considerations and numerical calculations, we can demonstrate that
the leading terms for $\langle \delta r\rangle$, $\langle \left(\delta r\right)
^2 \rangle$ and $\langle \left(\delta r\right)^3 \rangle$ are respectively
\begin{eqnarray}
\langle \delta r\rangle\propto {g_u}^{-1},~~\langle \left(\delta r\right)^2 \rangle\propto {g_u}^{-2},~~
\langle \left(\delta r\right)^3 \rangle\propto {g_u}^{-3}.
\end{eqnarray}
We substitute
\begin{eqnarray}
&&Z_1=r_\infty+\langle \delta r\rangle,~Z_2=r_\infty^2+2r_\infty\langle \delta r\rangle+\langle \left(\delta r\right)
^2 \rangle,\nonumber\\
&&Z_3= r_\infty^3+3r_\infty^2\langle \delta r\rangle+3r_\infty\langle \left(\delta r\right)^2 \rangle+\langle \left(\delta r\right)^3 \rangle,
\end{eqnarray}
into Eq.~(\ref{eq:iie2}) when $n=1$, 2 and 3 in the $l\rightarrow\infty$ limit
and obtain three coupled equations for $\langle \delta r\rangle$, $\langle \left(\delta r\right)^2 \rangle$
and $\langle \left(\delta r\right)^3 \rangle$.
We solve these equations analytically and substitute the resulting expressions for the three averages into Eq.~(\ref{eq:iie1})
to the leading order in the parameter ${g_u}^{-1}$.
The result for the localization length is given by
\begin{eqnarray}
k\xi=\frac{g_u}{{\beta_0}^2}=\frac{g_u}{\left(\sin\theta+a_0\right)^2},
\label{eq:ssp}
\end{eqnarray}
which is valid for all $\theta$ and diverges at $\theta_R$ given by Eq.~(\ref{eq:angle1}).
Close to $\theta_R$, it is approximated as
\begin{eqnarray}
k\xi\approx \frac{g_u}{\left(\cos^2\theta_R\right)\left(\theta-\theta_R\right)^2}.
\label{eq:z2}
\end{eqnarray}

We notice that the localization length does not depend on the average value of the scalar potential, $u_0$,
in the strong disorder regime.
We also notice that $\xi$ diverges in the infinite disorder limit for any values
of $u_0$, $a_0$ and $\theta$ since $k\xi\propto g_u$.
We have already predicted and explained this intriguing behavior using the impedance concept
in Sec.~\ref{sec:sdl}.
If there is no vector potential, then $k\xi$ is given by $k\xi=g_u/\sin^2\theta$ for all $\theta$.
We have verified that the agreement between Eq.~(\ref{eq:ssp}) and the numerical results
is perfect in the region where $g_u\gg 1$.

When only the vector potential is random in Model I, a similar derivation as above gives
the following expression for the localization length:
\begin{eqnarray}
\frac{1}{k\xi}=2\vert\beta_0\vert+\frac{{\epsilon_0}^2}{g_a}.
\label{eq:svp}
\end{eqnarray}
In the strong disorder limit, this expression approaches $2\vert\beta_0\vert$, which is independent of the disorder parameter.
Therefore, except at the angle where $\beta_0= 0$, the localization length is a finite constant independent of $g_a$.
This behavior has also been predicted using the impedance concept in Sec.~\ref{sec:sdl}.
Unfortunately, it turns out that the systematic truncation method we use for solving the invariant imbedding equations converges too slowly
in the present parameter regime and cannot provide sufficiently accurate results, and therefore we do not make comparisons between the IIM and
Eq.~(\ref{eq:svp}).

Finally, in the case of Model II, a similar strong-disorder perturbation method gives the following expression for the localization length:
\begin{eqnarray}
\frac{1}{k\xi}=\left\{\begin{matrix} \frac{1}{g_u}\left(\frac{\beta_0+f\epsilon_0}{1-f^2}\right)^2, &{\mbox {if}}~ \vert f\vert<1\cr
\frac{2\vert \epsilon_0 +f\beta_0\vert}{\sqrt{f^2-1}}+\frac{1}{g_u}\left(\frac{\beta_0+f\epsilon_0}{f^2-1}\right)^2, &{\mbox {if}}~ \vert f\vert>1\cr\end{matrix}.\right.
\label{eq:ssvp}
\end{eqnarray}
In the case where $\vert f\vert<1$, $k\xi$ diverges at the angle $\theta_S$ given by Eq.~(\ref{eq:angle2}).
In addition, in the infinite disorder limit, it diverges as $k\xi\propto g_u$ for all parameter values and for all $\theta$,
as in the case of Model I with only scalar potential disorder.
The agreement between Eq.~(\ref{eq:ssvp}) and the IIM results is perfect, as shown in Fig.~\ref{fig:comp}(c).
If $\vert f\vert>1$, the overall behavior is similar to Model I with only vector potential disorder.
As $g_u$ approaches infinity, $k\xi$ approaches an expression independent of $g_u$.
Therefore, except at the angle where $\beta_0+f\epsilon_0= 0$, the localization length is a finite constant independent of $g_u$.
The same behavior has also been predicted using the impedance concept in Sec.~\ref{sec:sdl}.
The truncation method used for solving our invariant imbedding equations converges too slowly
in the present parameter regime and cannot provide sufficiently accurate results.

\section{Numerical results}
\label{sec:res}

\begin{figure}
\centering\includegraphics[width=\linewidth]{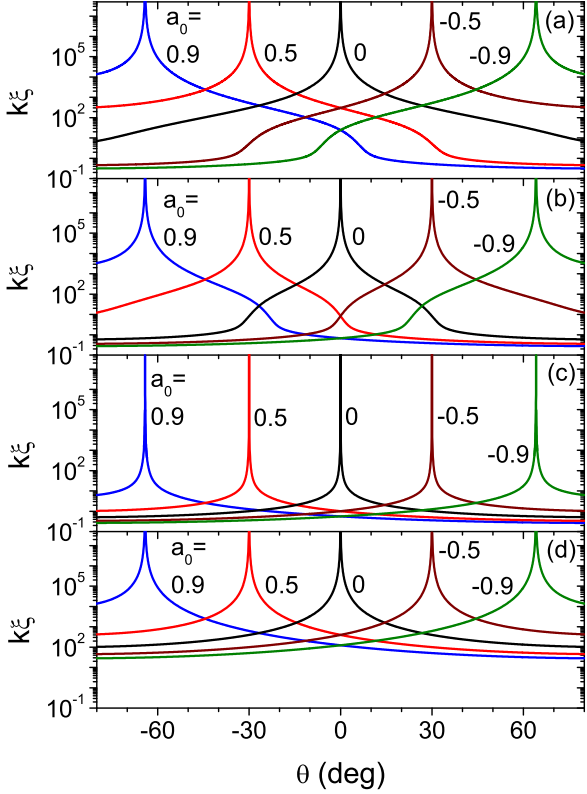}
 \caption{Normalized localization length $k\xi$ plotted versus incident angle $\theta$ for electron waves in Model I,
 when $g_a=0$, $a_0=0$, $\pm 0.5$, $\pm 0.9$ and (a) $g_u=0.01$, $u_0=0$, (b) $g_u=0.01$, $u_0=0.5$,
 (c) $g_u=0.01$, $u_0=1$, (d) $g_u=100$. There is no noticeable dependence on $u_0$ when $g_u=100$.}
 \label{fig:2}
 \end{figure}

\begin{figure}
\centering\includegraphics[width=\linewidth]{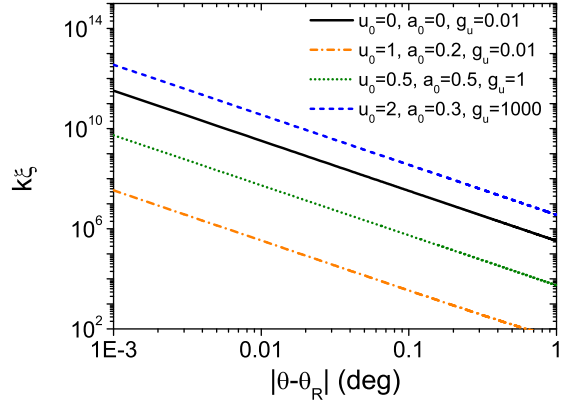}
 \caption{Normalized localization length $k\xi$ plotted versus $\vert\theta-\theta_R\vert$ in a log-log plot,
 for electron waves in Model I when only the scalar potential is random ($g_a=0$).
 The four curves are obtained for different values of the parameters $u_0$, $a_0$ and $g_u$, which are shown on the figure.
All curves show the divergent behavior $k\xi\propto \vert\theta-\theta_R\vert^{-2}$ near $\theta_R$.}
 \label{fig:th}
 \end{figure}

\begin{figure}
\centering\includegraphics[width=\linewidth]{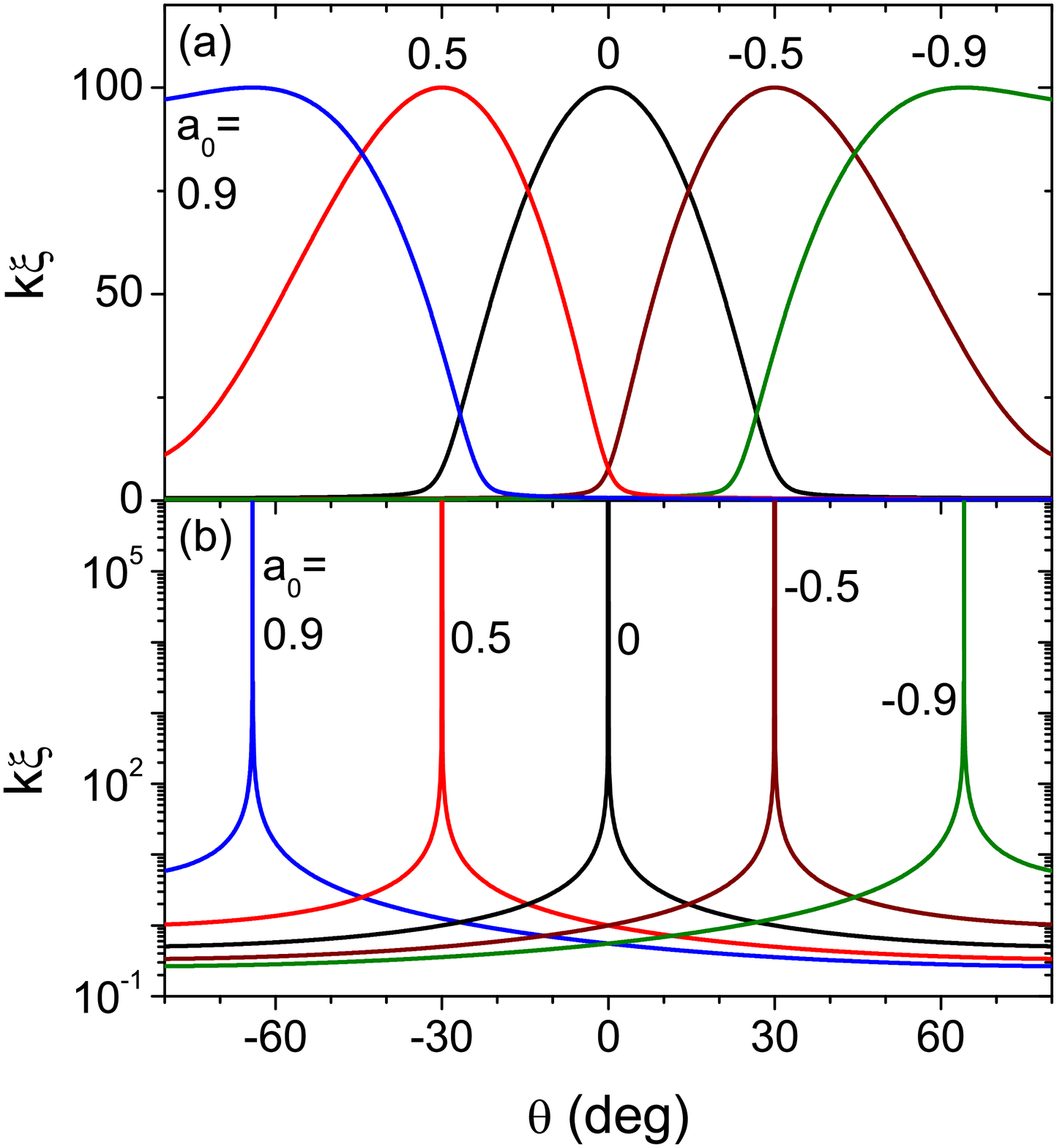}
 \caption{Normalized localization length $k\xi$ plotted versus incident angle $\theta$ for electron waves in Model I,
 when $g_a=0.01$, $g_u=0$, $a_0=0$, $\pm 0.5$, $\pm 0.9$ and (a) $u_0=0.5$, (b) $u_0=1$.}
 \label{fig:20}
 \end{figure}

\begin{figure}
\centering\includegraphics[width=\linewidth]{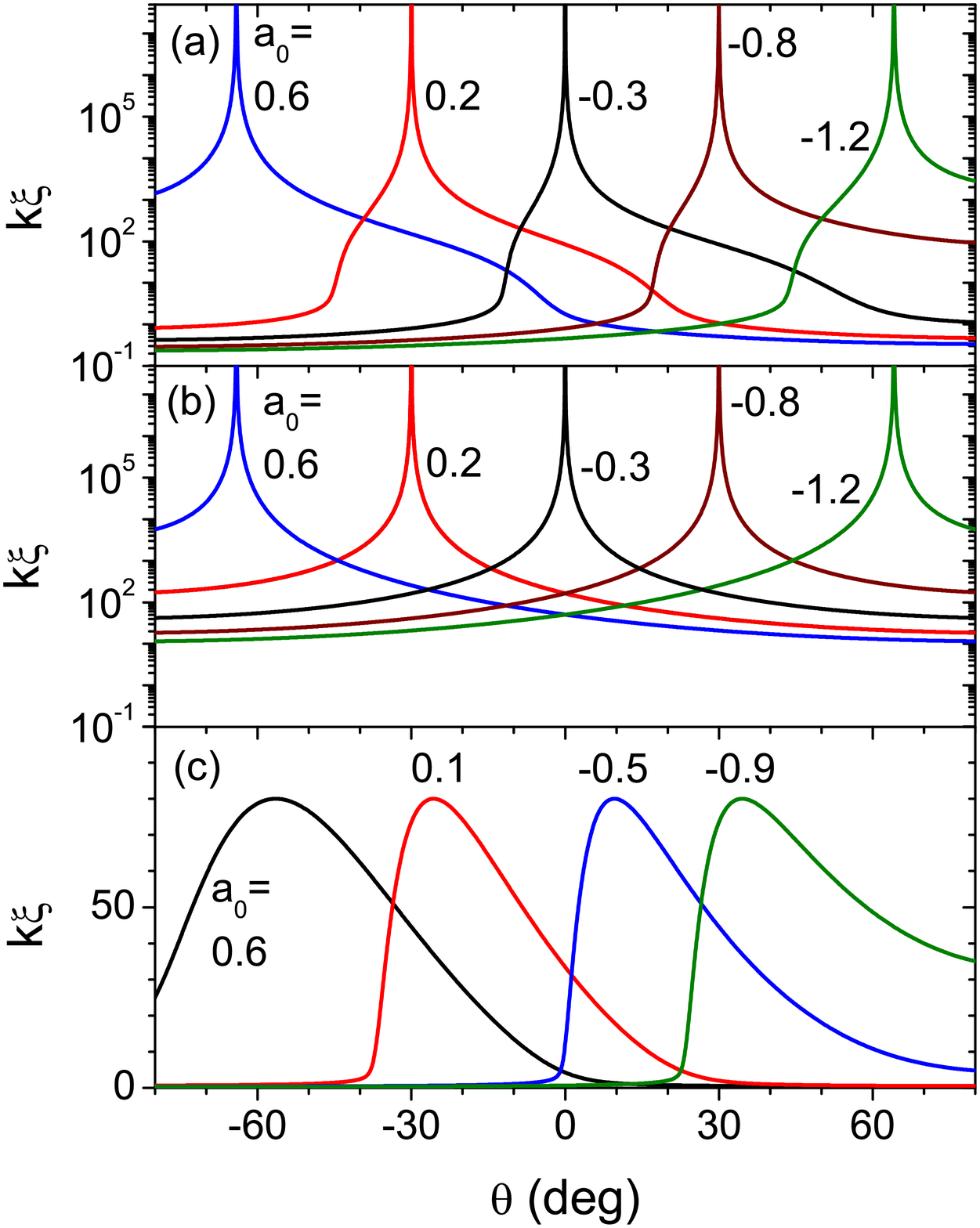}
 \caption{Normalized localization length $k\xi$ plotted versus incident angle $\theta$ for electron waves in Model II,
 when $u_0=0.5$. The values of $a_0$ are designated on the curves. The other parameters are (a) $f=0.6$, $g_u=0.01$,
 (b) $f=0.6$, $g_u=100$, and (c) $f=1.5$, $g_u=0.01$. }
 \label{fig:22}
 \end{figure}

\subsection{Incident angle dependence}

In this section, we present the results of our comprehensive numerical calculations obtained using the IIM.
We first consider the incident angle dependence of the localization length.
In Fig.~\ref{fig:2}, we plot the normalized localization length $k\xi$ as a function of the incident angle $\theta$
for electron waves in Model I, when only the scalar potential is random and the vector potential is constant.
We show the results for small ($g_u=0.01$) and large ($g_u=100$) values of the disorder parameter $g_u$.
When $g_u$ is small, we compare the results obtained for three different values of $u_0$ ($=0$, 0.5, 1). The dependence on $u_0$ disappears completely in the strong disorder regime.

In all cases, we find that the localization length indeed diverges precisely at $\theta_R$ given by $\sin\theta_R=-a_0$, if
$\vert a_0\vert\le 1$. If $\vert a_0\vert>1$, no Klein delocalization occurs. At $\theta_R$, the electron matter wave
is extended, as if it is in a nonrandom medium.
For $a_0=0$, $\pm 0.5$ and $\pm 0.9$ shown here, these angles are $0^\circ$, $\mp 30^\circ$ and $\mp 64.16^\circ$.
In addition, we find that the symmetry under the sign change of $\epsilon_0$ $(=1-u_0)$ and $\beta_0$ $(=\sin\theta+a_0)$ is strictly obeyed.
For instance, the results for $u_0=0$ and $u_0=2$ are identical and the curves for $a_0$ and $-a_0$ are mirror-symmetric
with respect to $\theta=0$. In fact, we have verified that the divergence of $\xi$ at $\theta_R$ and the symmetry property mentioned above are maintained
for all parameter values including the intermediate values of the disorder parameter.

The curves shown in Figs.~\ref{fig:2}(a) and \ref{fig:2}(b) agree extremely well with the analytical formula, Eq.~(\ref{eq:af1}),
except for very close to the critical angles of total reflection defined by $\sin\theta_c=-a_0\pm(1-u_0)$, if they exist.
Near $\theta_c$, a small difference in parameters can cause a large change of the reflection coefficient, and
therefore the perturbation theory is not reliable.
On the other hand, the curves shown in Fig.~\ref{fig:2}(d) agree perfectly with the formula in the strong disorder regime, Eq.~(\ref{eq:ssp}).

In Fig.~\ref{fig:2}(c), we show the weak disorder results obtained for $u_0=1$, which corresponds to the special case where $E=U_0$.
In this case, the expression $\beta/\epsilon$ in Eq.~(\ref{eq:imp1}) can be written as $-(\delta u/\beta_0)^{-1}$, where $\delta u/\beta_0$
plays the role of the effective random part. Therefore, as $\theta$ approaches $\theta_R$, this case becomes equivalent
to the case in the strong disorder limit where
Eq.~(\ref{eq:ssp}) is applied. In fact, this formula agrees with Fig.~\ref{fig:2}(c) quite well near $\theta_R$.

From the analytical formulas, Eqs.~(\ref{eq:z1}) and (\ref{eq:z2}), we find that in the presence of only scalar potential disorder,
the localization length shows a divergent behavior of the form $k\xi\propto \vert\theta-\theta_R\vert^{-2}$ near $\theta_R$
both in the weak and strong disorder limits.
In fact, this dependence is more general and applies also to
the intermediate disorder case. In Fig.~\ref{fig:th}, we illustrate this
point by showing the results for various parameter values in a log-log plot.

Next, we consider the case where only the vector potential is random in Model I.
In this case, our general argument based on the impedance concept and the analytical formulas for the localization length
show that the localization length is finite, except for the special case where both $\epsilon_0$ and $\beta_0$ are zero.
This is demonstrated in Fig.~\ref{fig:20}, where we show the results obtained for $g_a=0.01$, $g_u=0$ and $a_0=0$, $\pm 0.5$, $\pm 0.9$.
In Figs.~\ref{fig:20}(a) and \ref{fig:20}(b), the value of $u_0$ is 0.5 and 1 respectively. In Fig.~\ref{fig:20}(a), we confirm that
the localization length takes a maximum value given by ${g_a}^{-1}=100$ at the angle corresponding to $\beta_0=0$, in agreement with
Eq.~(\ref{eq:af1}). In Fig.~\ref{fig:20}(b), which corresponds to the case where $\epsilon_0=0$, we find that
the localization length indeed diverges at
the angle $\theta_R$. We have verified that the curves shown here agree very well with the formula, Eq.~(\ref{eq:z3}).

Finally, we consider the case of Model II, where the random parts of the vector and scalar potentials are proportional to each other
with the proportionality constant $f$. In this case, if $\vert f\vert<1$, the overall behavior is rather similar to Model I with
only scalar potential disorder, whereas if $\vert f\vert>1$, it is similar to Model I with
only vector potential disorder. We also notice that there is no symmetry with the sign change of
either $\epsilon_0$ or $\beta_0$. In Fig.~\ref{fig:22}(a), we show the weak disorder case where $\vert f\vert$ is smaller than 1.
We find that the localization length diverges precisely at the angle $\theta_S$ defined by Eq.~(\ref{eq:angle2}).
In Fig.~\ref{fig:22}(b), we show the strong disorder case where $\vert f\vert$ is smaller than 1.
Again, the localization length diverges at $\theta_S$ and the curves agree perfectly with Eq.~(\ref{eq:ssvp}).
In Fig.~\ref{fig:22}(c), we show the weak disorder case where $\vert f\vert$ is larger than 1.
Similarly to the case of Model I with only vector potential disorder, the localization length takes a finite maximum value
$[{g_u}(f^2-1)]^{-1}$ at the angle given by $\beta_0=-\epsilon_0/f$.

When the Klein delocalization occurs, our results show that the localization length has a single divergent peak at the Klein delocalization angle,
$\theta_R$ or $\theta_S$, depending on the model
and is finite elsewhere. In Ref.~37, a phenomenon termed delocalization resonance, referring to
the divergence or near-divergence of the localization length at angles other than the delocalization angle, was reported.
By generalizing our formalism to include a background periodic potential in addition to the random potential,
we have verified that this phenomenon occurs due to the background periodic potential and disappears completely if it is removed.
The complicated angle dependence of the transmittance showing disorder-induced resonances in Ref.~36 also occurs due to the interplay
between the background periodic potential and the random potential.

\begin{figure}
\centering\includegraphics[width=\linewidth]{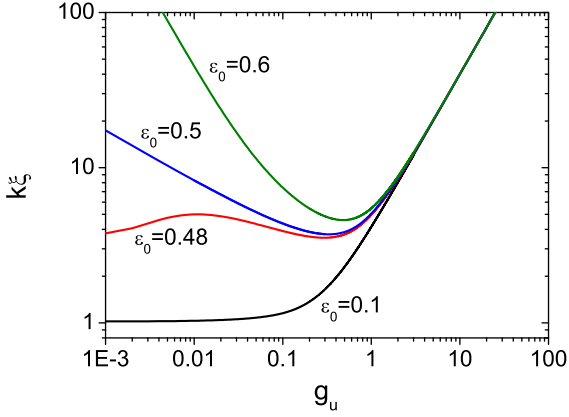}
 \caption{Normalized localization length plotted versus the strength of scalar potential disorder, $g_u$, for electron waves in Model I,
 when $g_a=0$, $\beta_0=0.5$ and $\epsilon_0=0.1$, 0.48, 0.5 and 0.6.}
 \label{fig:w1}
 \end{figure}

\begin{figure}
\centering\includegraphics[width=\linewidth]{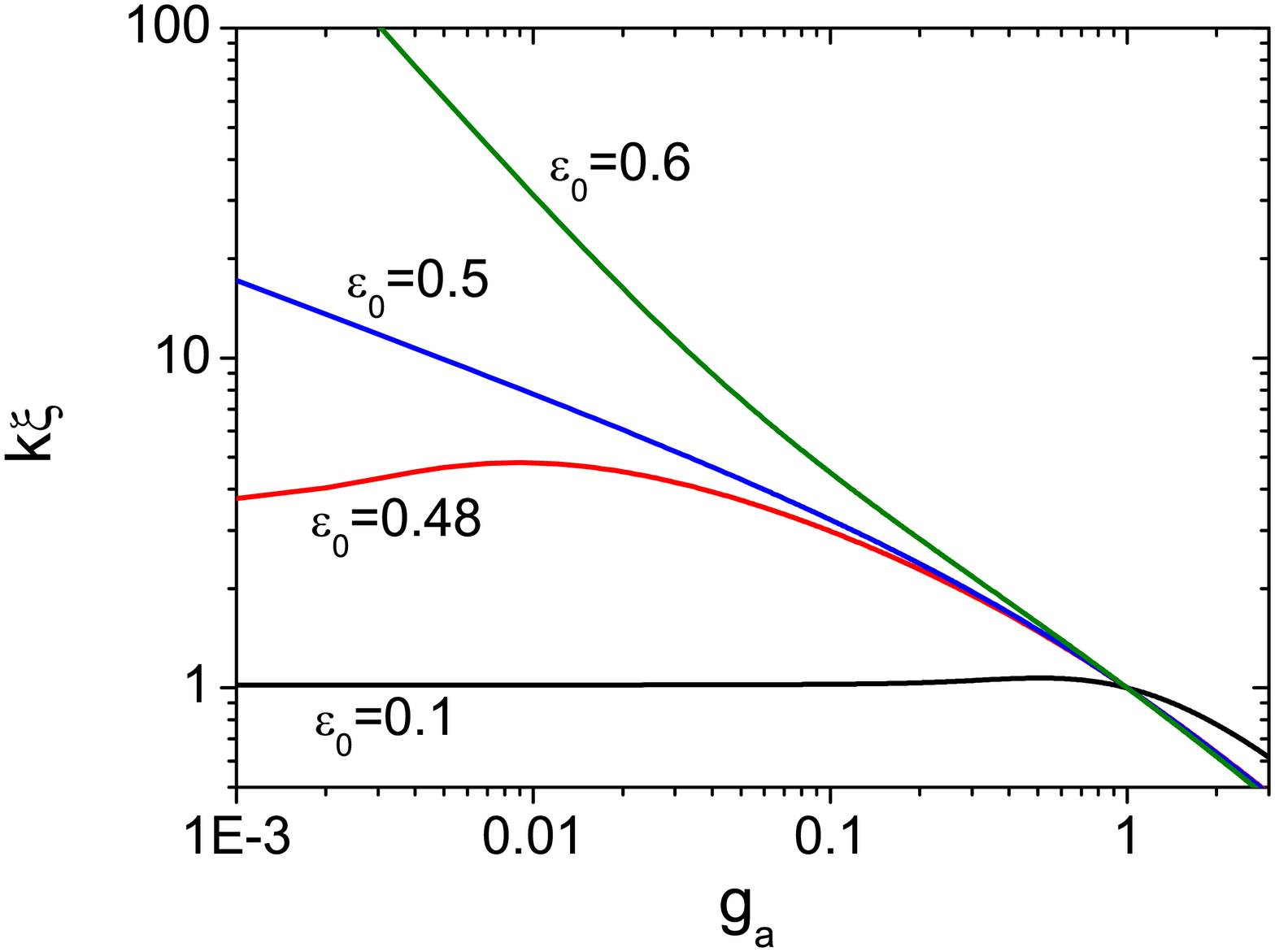}
 \caption{Normalized localization length plotted versus the strength of vector potential disorder, $g_a$, for electron waves in Model I,
 when $g_u=0$, $\beta_0=0.5$ and $\epsilon_0=0.1$, 0.48, 0.5 and 0.6.}
 \label{fig:w2}
 \end{figure}

\begin{figure}
\centering\includegraphics[width=\linewidth]{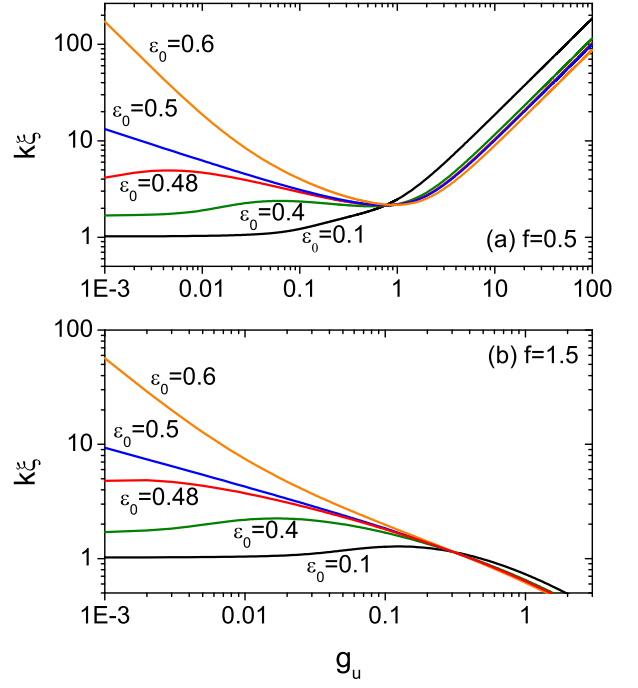}
 \caption{Normalized localization length plotted versus the strength of scalar potential disorder, $g_u$, for electron waves in Model II,
 when $\beta_0=0.5$, $\epsilon_0=0.1$, 0.4, 0.48, 0.5, 0.6 and (a) $f=0.5$, (b) $f=1.5$.}
 \label{fig:w3}
 \end{figure}

\subsection{Disorder dependence}
\label{sec:dd}

In this subsection, we consider the dependence of the localization length on the strength of disorder.
In Fig.~\ref{fig:w1}, we show the result for Model I with only scalar potential disorder, when $\beta_0$ is 0.5 and
$\epsilon_0$ takes values around $\beta_0$. When the vector potential is nonzero, the result
does not depend on $\theta$ and $a_0$ separately, but only on $\beta_0$ ($=\sin\theta+a_0$).
We find that there are three different types of behaviors, depending on the relative magnitudes of $\epsilon_0$ and $\beta_0$.
In the weak disorder regime, as the disorder parameter $g_u$ increases from zero,
the localization length is found to decrease if $\vert\epsilon_0\vert \ge \vert\beta_0\vert$ and increase if
$\vert\beta_0\vert>\vert\epsilon_0\vert$.
The latter behavior is a case of the disorder-enhanced tunneling phenomenon.
On the other hand, in the strong disorder regime, all curves converge to a single linear one given by $k\xi=g_u{\beta_0}^{-2}$,
which shows the divergence of $\xi$ in the $g_u\to\infty$ limit.
Therefore, in the region where $\vert\epsilon_0\vert \ge \vert\beta_0\vert$, the localization length has a nonmonotonic behavior
as shown by the curves for $\epsilon_0=0.6$ and 0.5.
On the contrary, in the opposite region where $\vert\epsilon_0\vert$ is sufficiently smaller than $\vert\beta_0\vert$,
the localization length increases monotonically as $g_u$ increases from zero to infinity,
as shown by the curve for $\epsilon_0=0.1$.
This type of behavior has never been reported before.
Sufficiently close to the boundary $\vert\epsilon_0\vert = \vert\beta_0\vert$ but when $\vert\epsilon_0\vert < \vert\beta_0\vert$,
we observe a very interesting behavior that $\xi$ increases initially, then decreases, and increases again to infinity,
as shown by the curve for $\epsilon_0=0.48$.
When $\beta_0=0.5$, we have found numerically that this third behavior occurs in the narrow region $0.44\lesssim\epsilon_0<0.5$.
This type of double-nonmonotonic behavior has also never been reported.
Our analytical formulas in Sec.~\ref{sec:form} cannot be used in the region where $\vert\epsilon_0\vert \approx \vert\beta_0\vert$.
Therefore the intriguing nonmonotonic behavior there can only be obtained numerically using the IIM.

Right at the boundary where $\vert\epsilon_0\vert = \vert\beta_0\vert$,
we observe that
the overall behavior is qualitatively similar to the case where $\vert\epsilon_0\vert > \vert\beta_0\vert$,
but the dependence of $\xi$ on $g_u$ in the small $g_u$ region is quantitatively different.
By careful numerical fitting of the data, we have found with very high accuracy that $\xi\propto {g_u}^{-1/3}$
when $\vert\epsilon_0\vert = \vert\beta_0\vert$, while $\xi\propto {g_u}^{-1}$
when $\vert\epsilon_0\vert > \vert\beta_0\vert$, in the small $g_u$ region. Contrasting scaling
behaviors of this kind are observed universally in the systems showing the disorder-enhanced tunneling phenomenon and are similar to
what happens to the electromagnetic waves incident at the critical angle on a randomly-stratified dielectric medium.\cite{kim7}

We next consider Model I with only vector potential disorder in Fig.~\ref{fig:w2}, when $\beta_0=0.5$.
The disorder-enhanced tunneling phenomenon also occurs in this case when $\vert\beta_0\vert>\vert\epsilon_0\vert$.
Therefore as the disorder parameter $g_a$ increases, $\xi$ initially increases if $\vert\beta_0\vert>\vert\epsilon_0\vert$
and decreases otherwise.
As $g_a$ increases further, $\xi$ is found to decrease in all cases.
In the region where $g_a$ is large, our numerical method converges very slowly and we cannot obtain reliable numerical results.
However, we expect from the argument based on the impedance concept and the analytical formula, Eq.~(\ref{eq:svp}), that
the localization length will approach a constant independent of $g_a$ in the $g_a\to\infty$ limit.

In Fig.~\ref{fig:w3}, we consider the case of Model II, when $\beta_0=0.5$.
If $\vert f\vert <1$ as in Fig.~\ref{fig:w3}(a), the overall behavior is quite similar to Model I with only scalar potential disorder.
One major difference from Fig.~\ref{fig:w1} is that in the strong disorder limit, the limiting behavior is precisely given by
$k\xi=g_u(1-f^2)^2/(\beta_0+f\epsilon_0)^2$, which depends on $\epsilon_0$ (therefore, $u_0$).
If $\vert f\vert >1$ as in Fig.~\ref{fig:w3}(b), the overall behavior is also quite similar to Model I with only vector potential disorder.
Again, in the large $g_u$ region, we cannot obtain reliable numerical results,
but we expect from the argument based on the impedance concept and the analytical formula, Eq.~(\ref{eq:ssvp}), that
the localization length will approach a constant independent of $g_u$ in the $g_u\to\infty$ limit.

\begin{figure}
\centering\includegraphics[width=\linewidth]{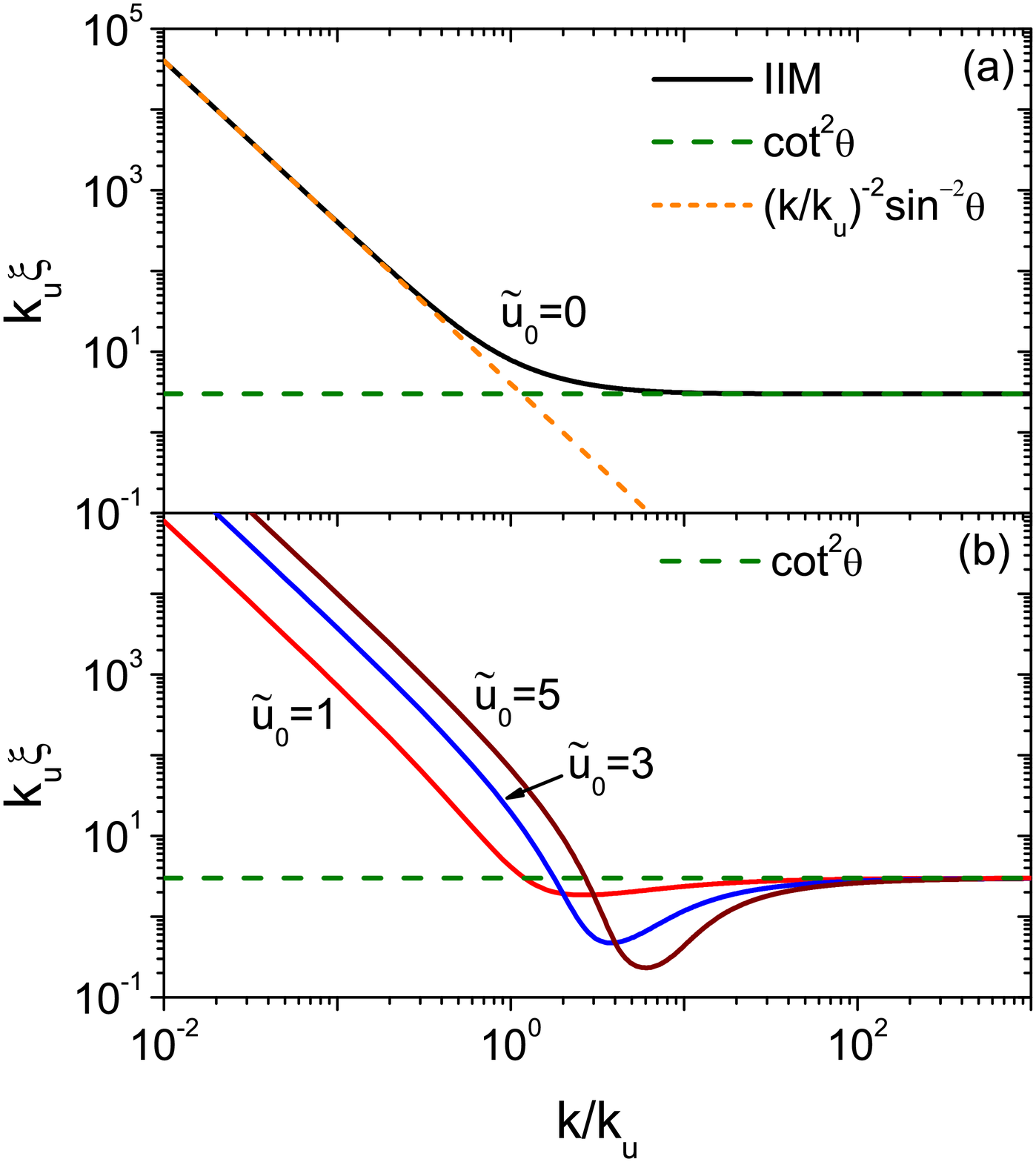}
 \caption{Energy dependence of the localization length in Model I when there is only a random scalar potential.
 $\xi$ is normalized by the wave number associated with disorder $k_u$ [see Eq.~(\ref{eq:disk0})] and $k/k_u$ [$=E/(\hbar v_F k_u)={g_u}^{-1}$] is the
 normalized energy variable. In (a), the parameter ${\tilde u}_0$ ($=u_0 k/k_u$) is zero and $\theta=30^\circ$. The IIM result is compared with
 the analytical results in the high and low energy limits. (b) ${\tilde u}_0=1$, 3, 5 and $\theta=30^\circ$.}
 \label{fig:w4}
 \end{figure}

\begin{figure}
\centering\includegraphics[width=\linewidth]{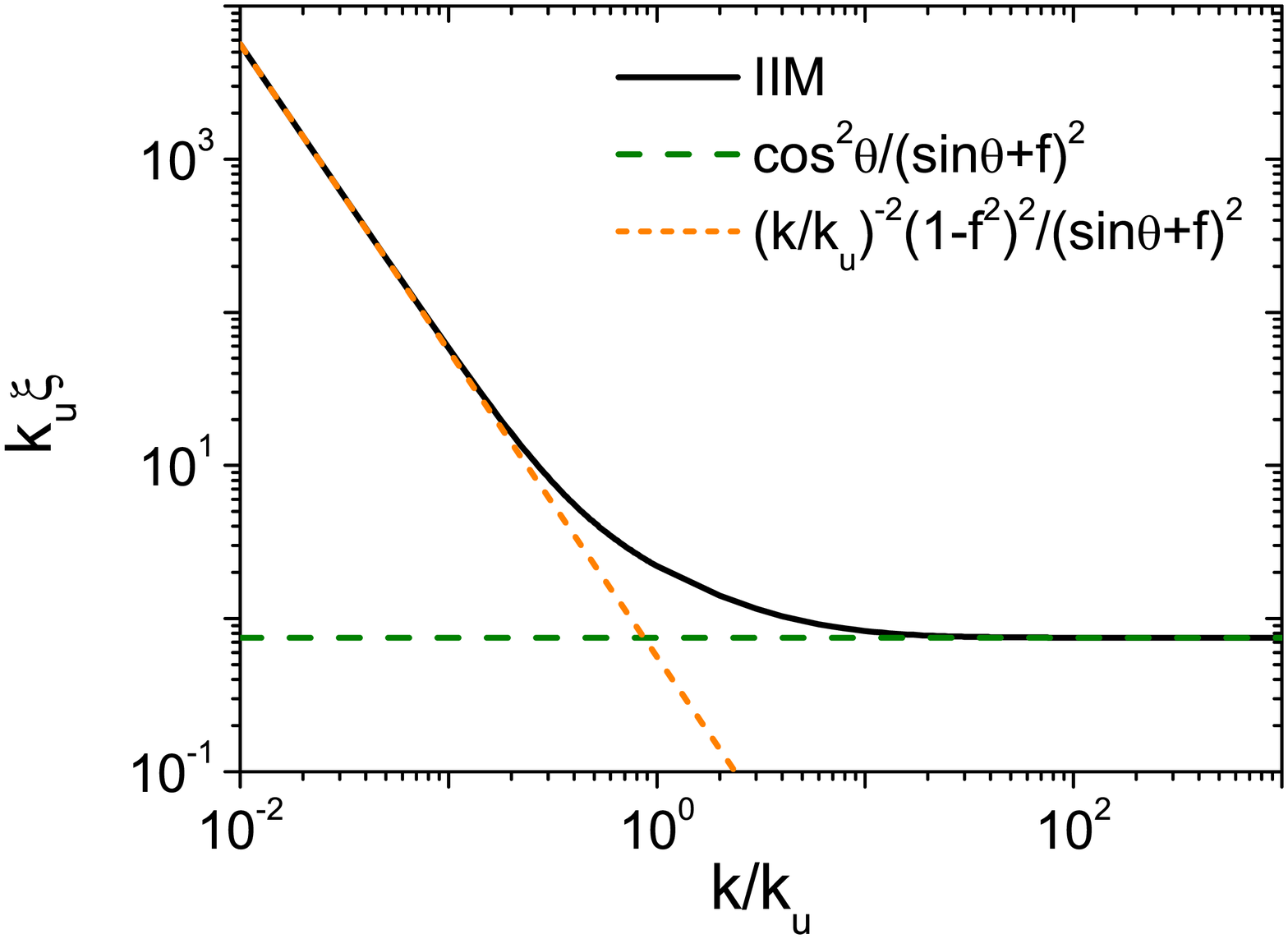}
 \caption{Energy dependence of the localization length in Model II with zero average values of the scalar and vector potentials
 when $\theta=30^\circ$. The IIM result is compared with
 the analytical results in the high and low energy limits. }
 \label{fig:w5}
 \end{figure}

\subsection{Energy dependence}

Until now, all physical quantities were made dimensionless by a suitable normalization using the energy of the incident particle, $E$, as
in $u=U/E$ and $a=eA_y/(\hbar ck)=ev_FA_y/(cE)$.
In order to obtain the energy dependence of the localization length properly, it is necessary
to redefine the dimensionless variables.
In this subsection, we will consider two cases, which are Model I with only scalar potential disorder and Model II with $\vert f\vert<1$.
For these cases, it is convenient to normalize all quantities using the wave number associated with the scalar potential disorder, $k_u$.
We suppose that the random part of the scalar potential $\delta U(x)$ satisfies
\begin{eqnarray}
\langle \delta U( x  )\delta U( x^\prime  )\rangle =G\delta(x-x^\prime).
\end{eqnarray}
Then $k_u$ is defined by
\begin{eqnarray}
k_u=\frac{G}{\left(\hbar v_F\right)^2}.
\label{eq:disk0}
\end{eqnarray}
It can be simply related to the variables we have used by
\begin{eqnarray}
k_u=k^2 {\tilde g}_u=kg_u.
\label{eq:disk}
\end{eqnarray}
We introduce new dimensionless variables
\begin{eqnarray}
{\tilde u}_0=\frac{\hbar v_FU_0}{G}=u_0\frac{k}{k_u},~~{\tilde a}_0=\frac{eA_{y0}}{\hbar  k_u}=a_0\frac{k}{k_u},
\end{eqnarray}
where $U_0$ and $A_{y0}$ are the disorder averages of the scalar and vector potentials
and $k/k_u$ [$=E/(\hbar v_F k_u)={g_u}^{-1}$] is the dimensionless energy variable.

In Fig.~\ref{fig:w4}, we plot the normalized localization length $k_u\xi$ versus $k/k_u$
in Model I with only scalar potential disorder and no vector potential,
for various values of ${\tilde u}_0$ and $\theta=30^\circ$.
In all cases, we find that in the low energy limit, $\xi$ is proportional to $E^{-2}$
and in the high energy limit, it is a constant independent of $E$.
The behavior in the high energy limit is surprising and somewhat counterintuitive in that the particles with extremely high energy
would show the same localization behavior as those with much lower energy.
We have also confirmed numerically that the limiting value of $k_u\xi$ in the high energy limit is universal and precisely the same as
$\cot^2\theta$, regardless of the average values of the scalar and vector potentials.
The low energy behavior $\xi\propto E^{-2}$ also always holds if ${\tilde a}_0$ is zero. However,
the proportionality constant depends on ${\tilde u}_0$.
We have found numerically that the low energy behavior is strongly modified in the presence of the vector
potential, but we will not discuss it here.
In Fig.~\ref{fig:w4}(b), we show that the localization length in the intermediate energy range,
where $k/k_u \approx {\tilde u}_0$ (or equivalently, $E\approx U_0$), is suppressed and
depends strongly on the potential.

The energy dependence obtained here is in a direct contradiction with that in Ref.~20 (see Eq.~15),
where it has been reported that in the large energy limit, the localization length has an oscillatory dependence on the energy
and its overall size is proportional to $E^2$. The oscillatory behavior is due to the background periodic potential
in their superlattice model. The reason for the discrepancy between their $E^2$ dependence and our constant result
is unclear at this stage and is worthy of further investigation. The energy dependence in the low energy region was not
reported in Ref.~20.
In Ref.~37, it has been stated that the localization length in the presence of a purely random scalar or vector potential
depends only on the incident angle and the disorder strength, but is independent of the energy.
This result agrees with our result in the high energy region, but is in contradiction with ours in the low energy region, where $\xi\propto E^{-2}$.
The weak-disorder expansion method used in Ref.~37 is not expected to be valid in the low energy region, where the effective
disorder is strong. Therefore the result of Ref.~37 cannot be applied to that region.

It is also interesting to compare our result with the case of the Schr\"odinger equation with a random scalar potential.
From the result obtained in Ref.~41, it is straightforward to deduce that
the localization length is proportional to $E$ in the high energy limit, while it goes to a constant
(with a possible logarithmic correction) in the low energy limit.

In Fig.~\ref{fig:w5}, we consider Model II with $\vert f\vert<1$.
We limit our interest to the case where the average potentials are zero.
Then we find that in the low energy limit, $\xi$ is proportional to $E^{-2}$
and in the high energy limit, it is a constant independent of $E$, similarly to the previous case.
The limiting behaviors agree precisely with those obtained analytically and shown on the figure.
In the presence of the nonzero average potentials, the behaviors are strongly modified, but we will not pursue that here.

\begin{figure}
\centering\includegraphics[width=\linewidth]{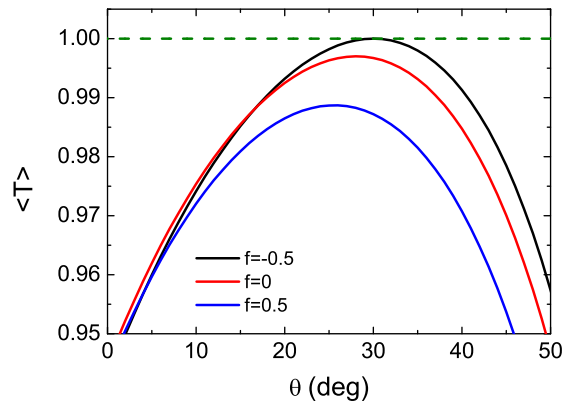}
 \caption{Disorder-averaged transmittance $\langle T\rangle$ versus incident angle
for Model II with $u_0=0.5$, $a_0=-0.25$, $g_u=0.01$ and $kL=1$ and for $f=-0.5$, 0, 0.5.}
 \label{fig:w6}
 \end{figure}

\subsection{Total transmission through a disordered region}

Finally, we briefly comment on the total transmission condition in the random case.
As we have stated in Sec.~\ref{sec:kl}, total transmission through finite random systems can arise
if $a(x)/u(x)$ remains a constant everywhere.
This can be achieved in our Model II with a suitable choice of $f$.
In order to verify this, we need to develop a method to calculate the disorder-averaged transmittance of a finite system.
We have done this using the IIM similar to the one developed in Sec.~\ref{sec:met}, which we do not elaborate here.
In Fig.~\ref{fig:w6}, we illustrate the total transmission phenomenon
by plotting the disorder-averaged transmittance $\langle T\rangle$ versus incident angle
for Model II with $u_0=0.5$, $a_0=-0.25$, $g_u=0.01$ and $kL=1$.
We find that when $f$ is equal to $a_0/u_0$ ($=-0.5$), $\langle T\rangle$ is indeed equal to 1 at the angle defined by Eq.~(\ref{eq:ktc}).

\section{Conclusion}
\label{sec:conc}

In this paper, we have studied Anderson localization and delocalization phenomena of Dirac electrons in 2D in 1D random scalar and vector potentials theoretically for two different cases. In the first case, the random parts of the scalar and vector potentials are uncorrelated while, in the second case, they are proportional to each other. We have calculated the localization length for all values of the disorder strength in a numerically exact manner using the IIM. We have also derived analytical expressions for the localization length, which are accurate in the weak and strong disorder regimes. We have generalized the condition for total transmission and those for delocalization to our random models and derived the incident angles at which obliquely incident electron waves are either completely transmitted or delocalized. We have found that these conditions, which include the ordinary Klein tunneling as a special case, are equivalent to the condition that the effective wave impedance is either matched or uniform. We have investigated the dependencies of the localization length on incident angle, disorder strength and particle energy in detail
and found crucial discrepancies with previous results and some surprising new results.

In addition to exploring novel localization phenomena which are completely different from those of nonrelativistic particles, our results have strong implications for electrical transport properties of graphene and similar 2D materials.
In Sec.~\ref{sec:kl}, we have proposed the use of inhomogeneous potentials for the design of very
efficient tunable directional filters. The use of random structures can also facilitate similar applications including
tunable electron beam supercollimation and other tunable electronic circuits.\cite{park,choich}
Our result can also be applied to the understanding of
wave propagation properties in equivalent photonic systems and to the design of photonic devices.

\acknowledgments
This work has been supported by the National Research Foundation of Korea Grant (NRF-2018R1D1A1B07042629) funded by the Korean Government.

\end{document}